# Excitation of unidirectional exchange spin waves by a nanoscale magnetic grating

Jilei Chen [1,†], Tao Yu [2,†], Chuanpu Liu [1,†], Tao Liu [3,†], Marco Madami [4,†], Ka Shen [5], Jianyu Zhang [1], Sa Tu [1], Md Shah Alam [1], Ke Xia [6], Mingzhong Wu [3], Gianluca Gubbiotti [7], Yaroslav M. Blanter [2], Gerrit E. W. Bauer [2,8,9,*] and Haiming Yu [1,*]

[1]*Fert Beijing Institute, BDBC, School of Microelectronics, Beihang University, Beijing, China.*

[2]*Kavli Institute of Nanoscience, Delft University of Technology, Delft, The Netherlands.*

[3]*Department of Physics, Colorado State University, Fort Collins, Colorado, USA.*

[4]*Dipartimento di Fisica e Geologia, Università di Perugia, Perugia, Italy.*

[5]*Department of Physics, Beijing Normal University, Beijing, China.*

[6]*Institute for Quantum Science and Engineering, Southern University of Science and Technology, Shenzhen, China.*

[7]*Istituto Officina dei Materiali del Consiglio Nazionale delle Ricerche (IOM-CNR), Sede di Perugia, c/o Dipartimento di Fisica e Geologia, Via A. Pascoli, I-06123 Perugia, Italy.*

[8]*Institute for Materials Research, WPI-AIMR and CSNR, Tohoku University, Sendai, Japan.*

[9]*Zernike Institute for Advanced Materials, University of Groningen, Groningen, The Netherlands.*

† These authors contributed equally to this work.

* g.e.bauer@tudelft.nl

haiming.yu@buaa.edu.cn

**Magnon spintronics is a prosperous field that promises beyond-CMOS technology based on elementary excitations of the magnetic order that act as information carriers for future computational architectures. Unidirectional propagation of spin waves is key to the realization of magnonic logic devices. However, previous efforts to enhance the Damon-Eshbach-type nonreciprocity did not realize (let alone control) purely unidirectional propagation. Here we experimentally demonstrate excitations of unidirectional exchange spin waves by a nanoscale magnetic grating consisting of Co nanowires fabricated on an ultrathin yttrium iron garnet film. We explain and model the nearly perfect unidirectional excitation by the chirality of the magneto-dipolar interactions between the Kittel mode of the nanowires and the exchange spin waves of the film. Reversal of the magnetic configurations of film and nanowire array from parallel to antiparallel changes the direction of the excited spin waves. Our results raise the prospect of a chiral magnonic logic without the need for fragile surface states.**

Spin waves (SWs)[1-6] can transport information in high-quality magnetic insulators such as yttrium iron garnet (YIG)[7-10] free of charge flow and with very low dissipation. Based on the interference and nonlinear interactions, the phase information of SWs[11-14] allows the design of wave-based logic circuits[15-17] for information transmission and processing with small environmental footprint. Surface SWs[18] are chiral, i.e. they propagate only in the direction of the outer product of magnetization direction and surface normal and, therefore, in opposite directions on the upper and lower film surfaces/interfaces. These "Damon-Eshbach" (DE) modes[18] are beneficial for magnonic logic devices[19] but exist only in thick magnetic films with sizable group velocities. As products of the dipolar magnetic interaction, in the case of thin films, they have small group velocities and are susceptible to surface roughness scattering. Previous efforts focused on magnetic metallic systems[20-28] with relative high dissipation. Short-wavelength spin waves[29-36] with dispersion governed by the exchange interactions, travel much faster at higher frequencies (Fig. 1**d**). However, pure exchange spin waves are not chiral, i.e. the travel equally well in all directions. Recently, Wintz et al.[29] observed spin waves in an intermediate regime with relatively short wavelength $\lambda = 125$ nm in small (4 μm) permalloy thin film structures that are "non-reciprocal", i.e. propagate with different velocities in opposite directions.

Here, we report unidirectional propagation of exchange spin waves (ESWs) down to wavelengths of 60 nm in ultrathin YIG films capped by an array of Co nanowires functioning as a nanoscale magnetic grating. The SW propagation direction can be controlled by changing the relative directions of the Co and YIG magnetizations from parallel (P-state) to antiparallel (AP-state). The chirality is strongly suppressed when the magnetizations are non-collinear. This property is important to realize either planar[30] or layered[31] reconfigurable magnonic crystals whose dynamic response can be controlled on demand by changing the magnetic configuration. Our observations cannot be explained by the excitation of the upper DE surface mode, because in ultrathin films the mode loses its surface character and acquires a quasi-uniform profile through the film thickness; the magnetization amplitudes of the fundamental SWs that propagate in both directions normal to the (in-plane) magnetization is practically identical[37]. Instead, we find that the interlayer dynamic dipolar coupling between the nanowires and the film is responsible for the observed effect. Our theoretical model suggests a nearly perfect unidirectional excitation of SWs when magnetizations are collinear and describes the angle dependence of the microwave spectra well. In addition, the unidirectional propagation of spin waves is further confirmed by micro-Brillouin light scattering (μ-BLS) spectroscopy.

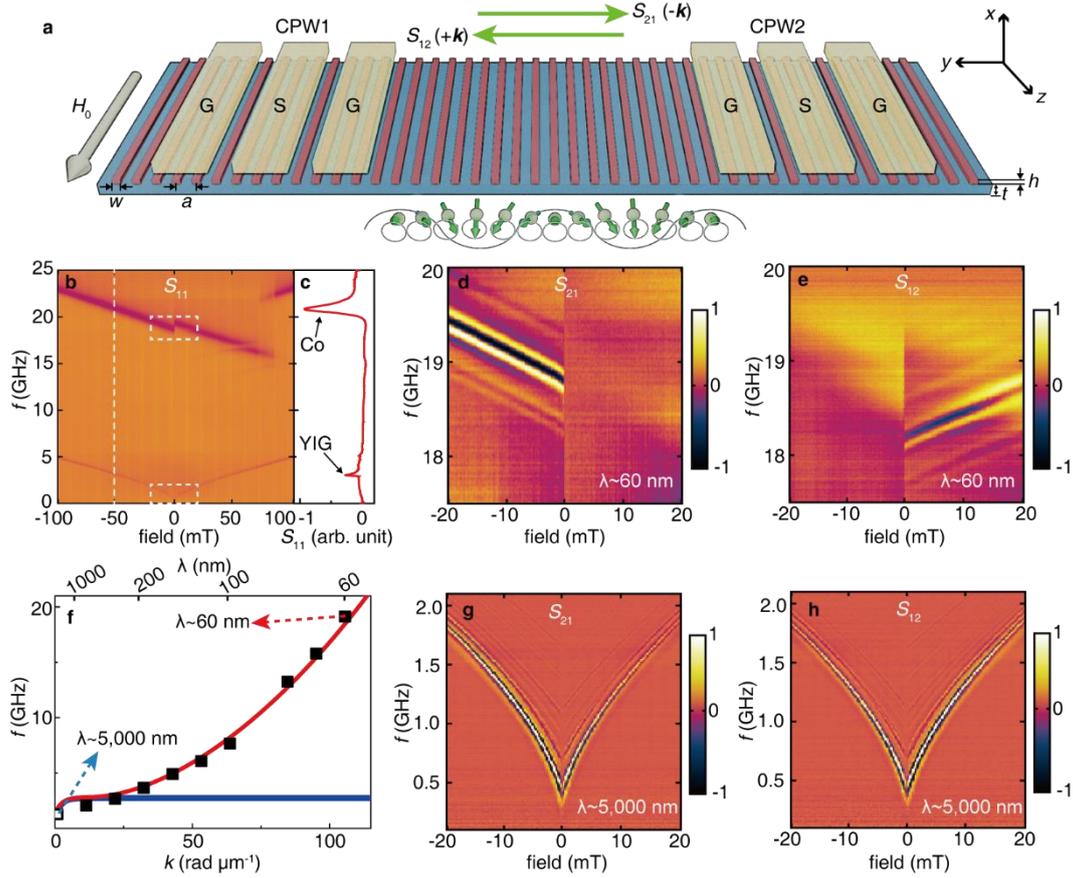

**Figure 1. Unidirectional propagation of short-wavelength exchange spin waves in Co/YIG magnetic nanostructures.**

**a,** Sketches of propagating spin waves in YIG thin film with a Co nanowire array on top. The YIG film is magnetically much softer than the Co nanowires. We consider here the P and AP configurations controlled by an external field $H_0$ applied in-plane and parallel to the nanowire length (easy magnetization direction). The SWs are excited by one CPW, propagate normal to the wires and are detected by the other CPW. The period of the nanowire array $a = 600$ nm and the width of a single nanowire is $w = 110$ nm. The high-frequency microwave transmission between the two waveguides mediated by SWs turns out to be unidirectional, i.e. when magnetizations are in P configuration ($H < 0$), $S_{21}$ is finite but $S_{21} = 0$. **b,** Microwave reflection spectra $S_{11}$ measured from -100 mT to 100 mT. A line plot is extracted from the white dashed line at -50 mT and shown in **c**. White dashed-line squares indicate the regions of exchange and dipolar SWs enlarged in panels **d-e** and **g-h**, respectively. **d,** The microwave transmission spectra $S_{21}$, carried by -**k** SWs with wave length λ from the left to the right at negative magnetic fields, but not positive ones. **e,** In contrast to **d** microwave transmission spectra $S_{12}$ carried by +**k** SWs in the opposite direction are transmitted only for positive fields. **f,** SW dispersion relation. The red curve represents the exchange-dipolar SW dispersion equation (1). The blue curve gives the dispersion of pure dipolar SWs (Damon-Eshbach modes). Black squares are data points extracted from the experiments (see SI for the extraction method). The highest mode number $n = 20$ corresponds to a propagating ESW with wavelength $\lambda \sim 60$ nm, while the lowest-frequency modes are dipolar SWs with wavelength $\lambda \sim 5,000$ nm. The low-frequency transmission spectra $S_{21}$ (**g**) and $S_{12}$ (**h**) are carried by (dipolar) spin waves degenerate with the YIG film Kittel mode resonance.

The sample and measurement set-up are sketched in Fig. 1**a**. We fabricated a periodic array of cobalt nanowires directly on top of an ultrathin YIG film grown on GGG substrate by magnetron sputtering[38] (see Methods) with period $a = 600$ nm and nanowire width $w = 110$ nm. The thickness of the YIG film $t = 20$ nm and the Co nanowires $h = 30$ nm. The tunable coplanar waveguides (CPWs) on top of the nanostructures excite and detect the magnetization dynamics (see Methods). The reflection spectrum $S_{11}$ measured at CPW1 (Fig. 1**b**) is almost identical to $S_{22}$ measured at CPW2 (see Fig. S1 in the SI). We first saturate the magnetization of YIG and Co nanowires with a large magnetic field of -200 mT along the nanowires and then sweep the field from negative to positive values. Figure 1**c** shows a lineplot extracted for -50 mT. The low (high) frequency mode is the ferromagnetic resonance (FMR) of the YIG film (Co nanowires), respectively. The soft magnetization of the YIG film switches at a very low field, while the Co nanowires have a much larger coercivity (~80 mT) due to their shape anisotropy[13,40]. Therefore, the Co/YIG bilayer assumes a stable antiparallel (AP) magnetic configuration from 0 to +80 mT as shown in Fig. 2**b**. We measured the delayed microwave propagation in both directions using a vector network analyzer (VNA)[11,12,14] over a distance of typically 15 μm. We plot the transmission spectra $S_{12}$ (+***k*** direction) and $S_{21}$ (-***k*** direction) in Figs. 1**d** and 1**e** for applied fields from -20 mT to 20 mT in the Damon-Eshbach (DE) configuration, i.e. parallel to the nanowires. For negative (positive) applied fields, we observe signal transmission by spin waves in the −*y* (+*y*) direction only, respectively.

At FMR frequencies around 1 GHz (low-frequency mode in Fig. 1b) the CPWs excite long-wavelength dipolar spin waves (DSWs) in YIG and show strong transmission signals in both $S_{12}$ (+$k$) and $S_{21}$ (-$k$), with weak DE-mode-induced nonreciprocity[18,37]. The periodic potential generated by the Co nanowire array in principle allows the excitation of higher spin-wave modes by a homogeneous microwave field, but our observation of short-wavelength modes in YIG at frequencies up to 19 GHz is unexpected. They become visible in the microwave transmission when nearly degenerate with the FMR of the Co nanowires for the following reason. The microwaves emitted by a CPW force the magnetization of the nanowires to precess in-phase. The latter generates a lattice-periodic dipolar field on the YIG magnetization with wave vectors $k = \frac{n\pi}{a}$ where $a$ is the nanowire period and $n = 2, 4, 6 ...$[32]. The perpendicular standing spin waves (PSSWs)[33,34] in films have mode numbers $m = 0, 1, 2, 3 ...$, but they are in the present thin film upshifted to above 35 GHz and can be disregarded. For large $n$ values, the SWs in the film are safely in the exchange-dominated regime with quadratic dispersion as shown in Fig. 1f. More precisely, these ESWs obey the dispersion relation[41] (see supplementary information for alternative derivation)

$$f = \frac{|\gamma|}{2\pi}\left[(H_0 + Ak^2)(H_0 + Ak^2 + \mu_0 M_S) + \left(1 - \frac{1-e^{-kt}}{kt}\right)\left(\frac{1-e^{-kt}}{kt}\right)\right]^{1/2}, \quad (1)$$

where $\gamma$ is the gyromagnetic ratio, $H_0$ is the applied field, $A = 4.88 \times 10^{-13}$ J m$^{-1}$ is the exchange stiffness constant and $\mu_0 M_S = 143.9$ kA/m is the saturation magnetization of sputtered YIG films[38]. The thickness of the YIG film $t = 20$ nm and $k$ is the (modulus of the) wave vector normal to the magnetization. We observe multiple of these ESW modes, which allows us to map the dispersion relation equation (1) in Fig. 1f (data shown in the SI). The microwave transmission spectra indicate a strong unidirectionality of the SW propagation, which can be controlled by a magnetic field. It changes sharply when the YIG magnetization switches from the parallel (P) to antiparallel (AP) configurations as shown in Figs. 2a and 2b with symmetry $S_{12}(M) = S_{21}(-M)$. This phenomenology coincides with that of chiral DE surface modes on thick magnetic slabs. However, here we can exclude this explanation: First, since the YIG film is so thin, there are no surface modes even in the dipolar regime at low frequencies, because the amplitudes of left and right moving spin waves are practically identical (Figs. 1**g** and 1**h**). Moreover, the short wave length modes are deep in the exchange regime at frequencies much higher than the DE limiting value (blue curve in Fig. 1**f**). We therefore conclude that the magnetic nanowires generate a unidirectionality that does not exist in the bare film.

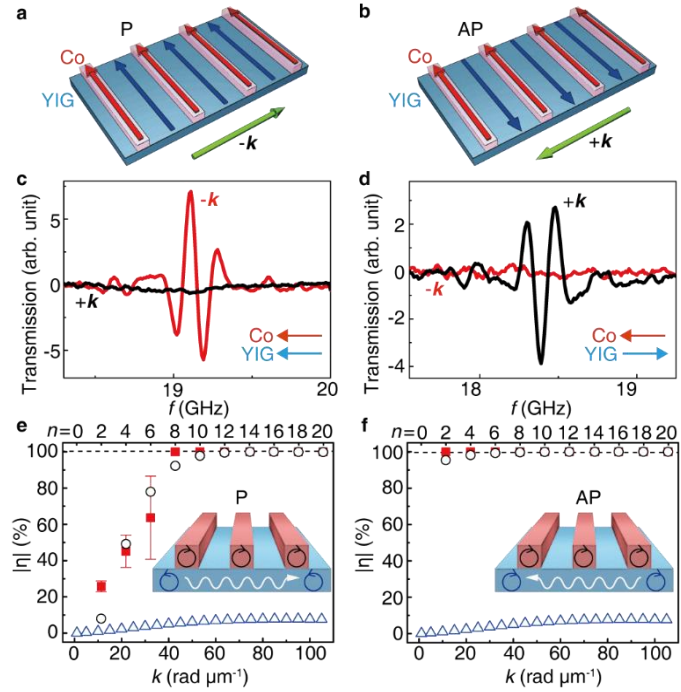

**Figure 2. Tunable chiral spin-wave excitations.** Co nanowires on top of a YIG films in the P (**a**) and AP (**b**) configurations. Microwave transmission by SWs is unidirectional and chiral, i.e. only -***k*** (+***k***) SWs can be excited in the P (AP) state, is illustrated by plotting $S_{21}^{(-k)}$ (red) and $S_{12}^{(+k)}$ (black) of the $n = 20$ ESW mode at -10 mT for the P (**c**) and AP state (**d**). We plot the degree of magnon chirality η, equation (2), as a function of the wave vector *k* for the P (**e**) and AP (**f**) configurations. The red filled squares represent the observations that agree well with micromagnetic theory (black open circles). The blue open triangles indicate the weak residual nonreciprocity in the low-frequency spin waves plotted in Fig. 1**g-h**.

Summarizing our observations, the Co/YIG bilayer in the P (AP) configuration only allows SW propagation in the -***k*** (+***k***) direction, appear to be chiral. We can quantify the magnon chirality in terms of the ratio

$$\eta = \frac{S_{21}^{(-k)} - S_{12}^{(+k)}}{S_{21}^{(-k)} + S_{12}^{(+k)}} , \qquad (2)$$

where $|\eta| = 1$ indicates perfect unidirectional spin-wave propagation. $|\eta|$ increases with the increasing mode number $n$ as shown in Figs. 2**e** and 2**f** for P and AP states, respectively. The exchange spin waves with $n \geq 8$ (cf. Fig. 1**d**) are fully chiral.

We confirmed the unidirectionality of SW propagation by micro-focused Brillouin light scattering (μ-BLS) spectroscopy excited by a coplanar antenna[39]. A single CPW antenna excites SWs in the specially designed sample sketched in Fig. 3**a**. Figure 3**b** shows the BLS spectra measured on both sides of the antenna under the same experimental conditions (P-state, $H = 20$ mT) as a function of the RF pumping signal. A spin signal is detected only on the right side of the antenna. By comparing BLS signals on the same (right) side for RF generator ON and OFF, we confirm that the thermal spin signal background is at resonance much smaller than the excited spin signal (Fig. 3**c**). With a reversed applied field (not shown) the signal is observed only on the left side.

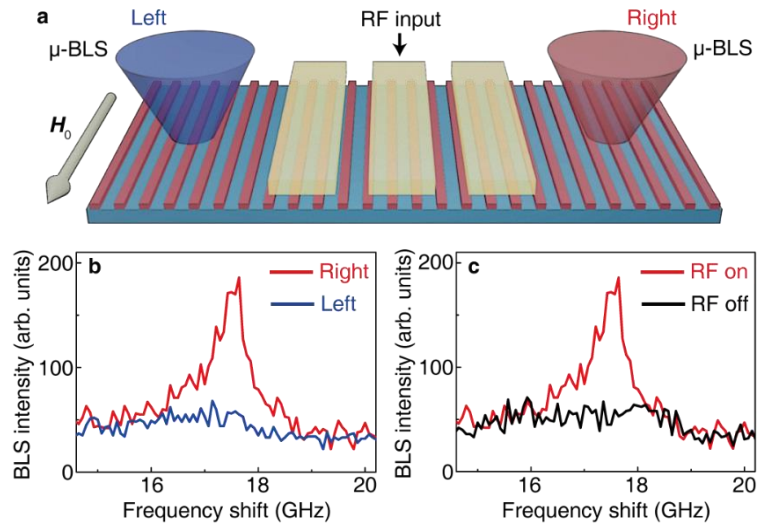

**Figure 3. Unidirectional propagation of exchange spin waves demonstrated by BLS measurements. a,** μ-BLS spectroscopy geometry on a Co nanowire array on YIG with parameters (defined in Fig. 1**a**) $w = 400$ nm, $a = 600$ nm and $h = 30$ nm. The applied external field ($H = 20$ mT) generates the P state and the laser focus spots are indicated. **b,** Comparison between two BLS spectra measured as a function of RF frequency on the right (red) and left (blue) sides of the antenna, respectively. **c,** Comparison between two BLS spectra measured on the same (right) side of the antenna with and without RF pumping.

We now turn to an explanation of the observed high magnon chirality of nominally non-chiral exchange spin waves. As discussed above, either the interface exchange or the dipolar interaction between the Co nanowires and YIG thin film is responsible for the effect. Since the former cannot generate the observed chirality, we focus here on the latter. Assuming perfect lattice and translational periodicity in the film plane and disregarding high frequency PSSWs, the micromagnetic problem becomes one-dimensional, with interaction Hamiltonian in second quantization

$$\widehat{\mathcal{H}}/\hbar = \sum_n (\omega_n^+ \hat{\beta}_{+k}\hat{a}^\dagger + \omega_n^- \hat{\beta}_{-k}\hat{a}^\dagger), \qquad (3)$$

where $\hat{a}^\dagger$ denotes the magnon creation operator for the Kittel mode of the Co nanowires, $\hat{\beta}_{+k}$ and $\hat{\beta}_{-k}$ are the magnon annihilation operators for the $+k$ and $-k$ spin waves of the YIG film, and

$$\begin{cases} \omega_n^+ = -\gamma\sigma_n \sqrt{(\mu_0 M_S^{Co})(\mu_0 M_S^{YIG})} \int \widehat{\boldsymbol{m}}_{Co}^* \widetilde{\boldsymbol{\Lambda}}^* \widehat{\boldsymbol{m}}_{YIG}\, e^{kx} dx \\ \omega_n^- = -\gamma\sigma_n \sqrt{(\mu_0 M_S^{Co})(\mu_0 M_S^{YIG})} \int \widehat{\boldsymbol{m}}_{Co}^* \widetilde{\boldsymbol{\Lambda}}\, \widehat{\boldsymbol{m}}_{YIG}\, e^{kx} dx \end{cases} \qquad (4)$$

is the coupling strength for spin waves with wave numbers $k = n\pi/a$ with $n = 2, 4, 6 \ldots$ propagating along the $+y$ direction for $\omega_n^+$ and $-y$ direction for $\omega_n^-$. Here, $M_S^{Co}$ and $M_S^{YIG}$ are the saturation magnetization of Co and YIG. In the P state, both magnetization are along $-z$ direction and the form factor $\sigma_n = \frac{2}{n\pi}\sin\left(\frac{kw}{2}\right)(1 - e^{-kh})$. $\widehat{\boldsymbol{m}}_{Co} = (\widehat{m}_x, \widehat{m}_y)$ is the Co magnetization procession of the nanowire Kittel mode, where $\widehat{m}_x = \left(\frac{a}{4hw}\sqrt{\frac{H_0 + M_S^{Co} N_{yy}}{H_0 + M_S^{Co} N_{xx}}}\right)^{1/2}$ and $\widehat{m}_y = \left(\frac{a}{4hw}\sqrt{\frac{H_0 + M_S^{Co} N_{xx}}{H_0 + M_S^{Co} N_{yy}}}\right)^{1/2}$, while $N_{xx}$ and $N_{yy}$ are the demagnetization factors[13,40] of a long wire. $\widetilde{\boldsymbol{\Lambda}} = \begin{pmatrix} 1 & i \\ i & -1 \end{pmatrix}$ and $\widehat{\boldsymbol{m}}_{YIG} = (\widehat{m}_x^k, \widehat{m}_y^k)$ describes the YIG magnetization procession of the SW modes. In the exchange regime, the spin precession is circular with $i\widehat{m}_x^k = \widehat{m}_y^k = i\left(\frac{1}{4t}\right)^{1/2}$. Derivations are given in the SI.

$|\omega_n^+| \neq |\omega_n^-|$ implies that the interlayer dynamic dipolar coupling between magnons in Co nanowires and YIG thin film is indeed chiral. We find for the magnon chirality factor equation (2)

$$\eta = \frac{(\omega_n^-/\omega_n^+)^2 - 1}{(\omega_n^-/\omega_n^+)^2 + 1}. \quad (5)$$

In the P configuration and the exchange regime, $\omega_n^- \neq 0$ and $\omega_n^+ = 0$, indicating perfect chirality or $\eta = 1$. For AP magnetizations, $\widehat{m}_{\text{YIG}} = (\widehat{m}_x^k, -\widehat{m}_y^k)$ yields $\omega_n^- = 0$ and $\omega_n^+ \neq 0$, indicating a reversed chirality compared with the P state as observed in perfect agreement with the analysis of the experimental data shown in Fig. 2**e** and Fig. 2**f**. The modelling results in Fig. 2**e** and Fig. 2**f** agree with experimental findings also for longer wavelength modes with partial chirality.

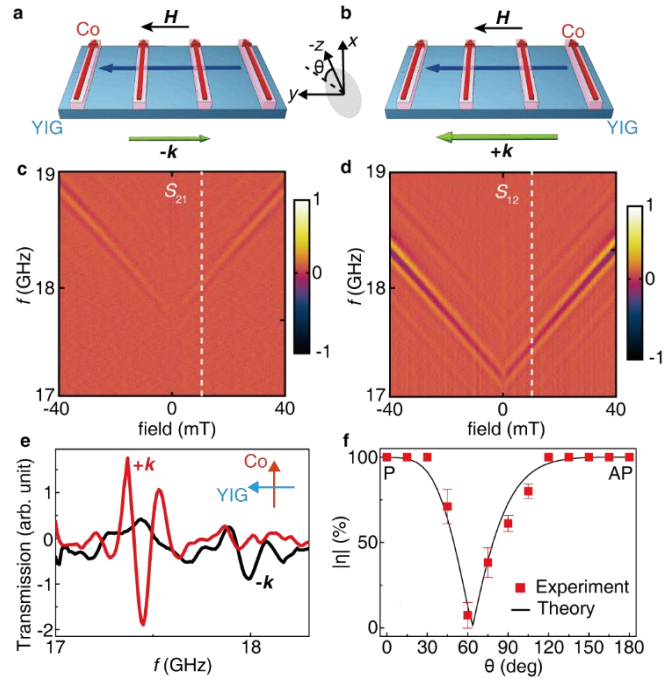

**Figure 4. Spin wave propagation in non-collinear configurations.** Microwave transmission by SWs with wave vector **-k** (**a**) and **+k** (**b**) when the film magnetization (blue arrow) is normal to the nanowires with magnetization indicated by the red arrows with (in-plane) angle $\theta = 90°$ (see inset). The SW transmission spectra $S_{21}$ for **-k** SWs (**c**) and $S_{12}$ for **+k** SWs (**d**) as function of an applied field $H$ along the $y$ axis after saturating the nanowire magnetization with a -200 mT field along $z$. (**e**) Transmission spectra for $\theta = 90°$ at 10 mT for $S_{21}$ (**-k**, black) and $S_{12}$ (**+k**, red) for $n = 20$ as indicated by the white dashed lines in **c** and **d**. The inset illustrates the magnetization directions. (**f**) $\theta$ dependence of the magnon chirality $\eta$ as observed (red squares) and calculated (black line). $\theta = 0°$ and $180°$ indicate the P and AP states, respectively.

We can coerce the magnetizations of the Co nanowires and the YIG thin film to form a finite angle θ by relatively weak external magnetic fields (see Fig. 4**a** and **b** for θ = 90°) because the YIG coercivity is very small[38]. The transmission spectra were recorded after saturating the nanowire magnetizations with an applied field of -200 mT in the $z$ direction. The magnetization of the YIG thin film is saturated already at a relatively small fields (< 50 mT) along $y$, but does not affect the magnetization of the Co nanowires along $z$ axis due to their large demagnetization field (~80 mT). We observe in Figs. 4**c** and 4**d** that at θ = 90° the chirality is broken. i.e. SWs propagate in both +***k*** and –***k*** directions. A finite non-reciprocity persists in the spectra at a field of 10 mT in Fig. 4**e**, different from a DE-mode-induced nonreciprocity that would have been transformed into fully reciprocal backward moving volume modes. We show the magnon chirality |η| of the $n = 20$ ESW extracted from the experiments as a function of θ in Fig. 4**f**. The micromagnetic model equations (4) and (5) (solid black curve) reproduces the observed angular dependence of the chirality (red squares) very well (see SI for details). The frequency shift $\delta f \sim 0.5$ GHz between +***k*** and –***k*** spin waves in Fig. 4**e** is a feature beyond the interlayer dipolar coupling model. The chirality persists in another sample with an $Al_2O_3$ spacer between Co and YIG (see Fig. S3 in the SI) indicating that the interlayer dynamic dipolar coupling is the key mechanism for the observed chirality. On the other hand, the insulating barrier quenches the frequency shift with θ, which we therefore associate with the Co-YIG exchange coupling[32,34,42] and an associated exchange-spring magnetization texture or an interfacial Dzyaloshinskii-Moriya interaction[43].

The interaction between the magnetic modes in the nanowire array and the continuous film is strong, i.e. total coupling strength $\omega_n = |\omega_n^+| + |\omega_n^-|$ can be measured in terms of the anticrossing gap between the Kittel mode of Co nanowires and the YIG SW modes[32,34,42].

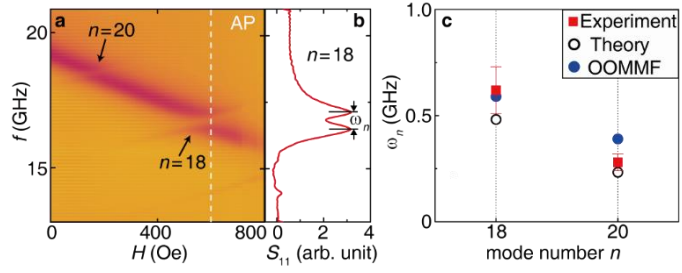

**Figure 5. Total coupling strength characterized by the anticrossing gaps. a,** SW reflection spectra $S_{11}$ measured in the AP state. Arrows indicate two observed anticrossings for $n = 18$ and $n = 20$ ESWs, respectively. **b,** A single SW spectrum measured at 60 mT which allows extraction of the anticrossing gap $\omega_n^+$ of the $n = 18$ mode. **c,** Anticrossing gaps for two SW modes: Red squares are experimental data; black circles are derived from the theoretical model; blue dots are the results of micromagnetic simulations with the OOMMF code[45].

Figure 5a shows SW reflection spectra measured in the AP state. Two clear anticrossings indicated couplings between the Kittel mode of the Co nanowires and two ESWs in YIG. The anticrossing for the mode $n = 16$ (see SI for the full spectra) is only partially resolved because the AP state switches back to the P state at fields of ~80 mT that overcome the demagnetizing field of the Co nanowires. The interaction splits the resonance peak at $n = 18$ of the lineplot in Fig. 5b by $\omega_n \cong 0.62$ GHz. This anticrossing also occurs when a magnetic field of 60 mT is rotated in the film plane (see SI). From the observed line widths, we extract a dissipation rate $\kappa_m^{Co} \cong 0.81$ GHz for the Kittel mode in Co and $\kappa_m^{YIG} \cong 0.11$ GHz for the exchange spin waves in YIG. Since $\kappa_m^{YIG} < \omega_n < \kappa_m^{Co}$ we are in the magnetically induced transparency, but not the strong coupling regime[44]. We plot the computed purely dipolar coupling using equation (4) in Fig. 5c as open circles. Since the magnon chirality rate $|\eta| \approx$ 100% for high-order exchange spin waves (Fig. 2f), $\omega_n^+ \approx \omega_n$ and $\omega_n^- \approx 0$ for the AP states. Micromagnetic simulations based on the object oriented micromagnetic framework (OOMMF)[45] confirm the anti-crossing features (see SI for the details of the simulations) and the computed gaps are the blue dots in Fig. 5c.

In conclusion, we report generation of unidirectional beams of non-chiral exchange spin waves in YIG film with wavelengths down to 60 nm when brought to resonance with a Co nanowire grating coupler[46-48] on top of the film. The spin-wave direction can be reversed by switching the magnetic configuration of Co nanowire array and YIG film from parallel to antiparallel. We explain and model the experimental results by the interlayer dipolar coupling between Co and YIG that can be assessed by the anticrossing gaps at the resonance. We observe and model a nearly perfect chirality for magnon propagation in a collinear magnetic configuration. Non-collinear applied magnetic field break the symmetry and suppress chirality. Interestingly, our findings appear to be a magnonic counterpart of the unidirectional excitation of surface plasmon polariton waves by circularly polarized electric dipoles[49], but note remarkable differences as well. The ability to excite unidirectional and easily switchable exchange spin waves with high group velocity can become a key functionality in reconfigurable nano-magnonic logic and computing devices[30,6,50].

**Methods**

**Sample fabrication.** The 20-nm-thick YIG thin film is deposited on a 0.5-mm-thick (111) oriented gadolinium gallium garnet (GGG) substrate by the sputtering deposition. The deposition is at room temperature first and the annealing is in $O_2$ at 800 °C. A YIG waveguide with the dimensions of 250 μm × 90 μm is fabricated by the photolithography and the ion beam etching. The YIG waveguide is tapered at both ends to avoid the spin wave interference. An array of Co nanowires is patterned on top of the YIG waveguide by electron beam lithography and electron beam evaporation. The width and the period of the Co nanowires are 110 nm and 600 nm respectively. A pair of identical gold coplanar waveguides (CPW) are then patterned on top of the Co nanowires with the signal line and ground line width of 2 μm. The gap between the signal line and the ground line is 1.6 μm. The center-to-center distance between two CPWs is 15 μm.

**All electrical spin-wave measurement.** The spin-wave measurement is conducted by a vector network analyzer (VNA) based all electrical technique. The Rohde & Schwarz ZVA 40 VNA used in the experiment offers a frequency range from 10 MHz to 40 GHz. The CPWs are connected with the

GGB MODEL 40A microwave probes. The power used in the experiment is 0 dBm. During the measurement, the VNA sends a microwave current to the CPW and generates an alternating magnetic field which excites the spin waves in the ferromagnetic film. The spin waves induced magnetic flux change is also detected by the VNA in reverse. The *S*-parameters with both reflection ($S_{11}$ and $S_{22}$) and transmission ($S_{12}$ and $S_{21}$) spectra are extracted from the VNA.

**Micro-BLS measurement.** The μ-BLS measurements were performed by scanning a diffraction limited laser beam, with a diameter of about 235 nm[39], on both sides of the CPW antenna and along the direction of propagating spin waves (perpendicular to the direction of *H*). The backscattered light is analysed in frequency by a (3+3)-tandem Sandercock-type Fabry-Pérot interferometer. To excite spin waves the CPW antenna was connected via RF picoprobes to a microwave generator working up to 20 GHz and with +20 dBm microwave power. Sample stabilization against mechanical drifts was achieved by the TFPDAS4 software which employs both an image recognition and an auto-focusing routine to compensate, in real time, for drifts of the sample position and of the focus distance.

**Acknowledgement**

We thank K. Buchanan for helpful discussions. We wish to acknowledge the support by NSF China under Grant Nos. 11674020, U1801661 and 111 talent program B16001. T.Y. and Y.B. was supported by the Netherlands Organization for Scientific Research (NWO). G.B. was supported by Japan Society for the Promotion of Science Kakenhi (Japan) Grants-in-Aid for Scientific Research (Grant No. 26103006). K.X. thanks the National Key Research and Development Program of China (2017YFA0303304, 2018YFB0407601) and the National Natural Science Foundation of China (61774017, 11734004). T.L. and M.W. were supported by the U.S. National Science Foundation (EFMA-1641989) and the U.S. Department of Energy, Office of Science, Basic Energy Sciences (DE-SC0018994).

**References**

1   Chumak, A. V., Vasyuchka, V. I., Serga, A. A. & Hillebrands, B. Magnon spintronics. *Nat. Phys.* **11**, 453-461 (2015).
2   Kruglyak, V. V., Demokrotiv, S. O. & Grundler, D. Magnonics. *J. Phys. D: Appl. Phys.* **43**, 264001 (2010).
3   Demidov, V. E. *et al.* Magnetization oscillations and waves driven by pure spin currents. *Phys. Rep.* **673**, 1-23 (2017).
4   Yu, H., Xiao, J. & Pirro, P. Magnon spintronics. *J. Magn. Magn. Mater.* **450**, 1-2 (2018).

5  Haldar, A., Kumar, D. & Adeyeye, A. O. A reconfigurable waveguide for energy-efficient transmission and local manipulation of information in a nanomagnetic device. *Nat. Nanotechnol.* **11**, 437-443 (2016).

6  Wagner, K. *et al.* Magnetic domain walls as reconfigurable spin-wave nanochannels. *Nat. Nanotechnol.* **11**, 432-436 (2016).

7  Serga, A. A., Chumak, A. V. & Hillebrands, B. YIG magnonics. *J. Phys. D: Appl. Phys.* **43**, 264002 (2010).

8  Kajiwara, Y. *et al.* Transmission of electrical signals by spin-wave interconversion in a magnetic insulator. *Nature* **464**, 262–266 (2010).

9  Chang, H. *et al.* Nanometer-thick yttrium iron garnet films with extremely low damping. *IEEE Magn. Lett.* **5**, 6700104 (2014).

10 Cornelissen, L. J., Liu, J., Duine, R. A., Youssef, J. B. & van Wees, B. J. Long-distance transport of magnon spin information in a magnetic insulator at room temperature. *Nat. Phys.* **11**, 1022–1026 (2015).

11 Vlaminck, V. & Bailleul, M. Current-induced spin-wave Doppler shift. *Science* **322**, 410–413 (2008).

12 Neusser, S. *et al.* Anisotropic propagation and damping of spin waves in a nanopatterned antidot lattice. *Phys. Rev. Lett.* **105**, 067208 (2010).

13 Ding, J., Kostylev, M. P. & Adeyeye, A. O., Magnonic crystal as a medium with tunable disorder on a periodical lattice. *Phys. Rev. Lett.* **107**, 047205 (2011).

14 Yu, H. *et al.* Magnetic thin-film insulator with ultra-low spin wave damping for coherent nanomagnonics. *Sci. Rep.* **4**, 6848 (2014).

15 Khitun, A., Bao, M. & Wang, K. L. Magnonic logic circuits. *J. Phys. D: Appl. Phys.* 43, 264005 (2010).

16 Schneider, T. *et al.* Realization of spin-wave logic gates. *Appl. Phys. Lett.* **92**, 022505 (2008).

17 Csaba, G., Papp, A. & Porod, W. Perspectives of using spin waves for computing and signal processing. *Phys. Lett. A* **381**, 1471-1476 (2017).

18 Damon, R. W. & Eshbach, J. R. Magnetostatic modes of a ferromagnet slab. *J. Phys. Chem. Solids* **19**, 308 (1961).

19 Jamali, M., Kwon, J. H., Seo, S.-M., Lee, K.-J. & Yang, H. Spin wave nonreciprocity for logic device applications. *Sci. Rep.* **3**, 3160 (2013).

20 Khalili-Amiri, P., Rejaei, B., Vroubel, M. & Zhuang, Y. Nonreciprocal spin wave spectroscopy of thin Ni–Fe stripes. *Appl. Phys. Lett.* **91**, 062502 (2007).

21 Schneider, T., Serga, A. A., Neumann, T., Hillebrands, B. & Kostylev, M. P. Phase reciprocity of spin-wave excitation by a microstrip antenna. *Phys. Rev. B* **77**, 214411 (2008).

22 Demidov, V. E., Kostylev, M. P., Rott, K., Krzysteczko, P., Reiss G. & Demokritov S. O. Excitation of microwaveguide modes by a stripe antenna. *Appl. Phys. Lett.* **95**, 112509 (2009).

23 Sekiguchi, K. *et al.* Nonreciprocal emission of spin-wave packet in FeNi film. *Appl. Phys. Lett.* **97**, 022508 (2010).

24 Kostylev, M. Non-reciprocity of dipole-exchange spin waves in thin ferromagnetic films. *J. Appl. Phys.* **113**, 053907 (2013).

25 Mruczkiewicz, M. *et al.* Nonreciprocity of spin waves in metallized magnonic crystal. *New J. Phys.* **15**, 113023 (2013).

26 Di, K. *et al.* Enhancement of spin-wave nonreciprocity in magnonic crystals via synthetic antiferromagnetic coupling. *Sci. Rep.* **5**, 10153 (2015).

27 Kwon, J. H. *et al.* Giant nonreciprocal emission of spin waves in Ta/Py bilayers. *Sci. Adv.* **2**, 1501892 (2016).

28  Mruczkiewicz, M. *et al.* Spin wave nonreciprocity and magnonic band structure in thin permalloy film induced by dynamical coupling with an array of Ni stripes, *Phys Rev. B* **96**, 104411 (2017).

29  Wintz, S. *et al.* Magnetic vortex cores as tunable spin-wave emitters. *Nat. Nanotechnol.* **11**, 948–953 (2016).

30  Grundler, D. Reconfigurable magnonics heats up. *Nat. Phys.* **11**, 438-441 (2015).

31  Gubbiotti, G., Zhou, X., Haghshenasfard, Z., Cottam, M. G. & Adeyeye, A. O. Reprogrammable magnonic band structure of layered Permalloy/Cu/Permalloy nanowires. *Phys. Rev. B* **97**, 134428 (2018).

32  Chen, J. *et al.* Strong interlayer magnon-magnon coupling in magnetic metal-insulator hybrid nanostructures. *Phys. Rev. Lett.* **120**, 217202 (2018).

33  Kittel, C. Excitation of spin waves in a ferromagnet by a uniform rf field. *Phys. Rev.* **110**, 1295-1297 (1958).

34  Qin, H., Hämäläinen, S. J. & van Dijken, S. Exchange-torque-induced excitation of perpendicular standing spin waves in nanometer-thick YIG films. *Sci. Rep.* **8**, 5755 (2018).

35  Liu, C. *et al.* Long-distance propagation of short-wavelength spin waves. *Nat. Commun.* **9**, 738 (2018).

36  Hämäläinen, S. J., Brandl, F., Franke, K. J. A., Grundler, D. & van Dijken, S. Tunable short-wavelength spin-wave emission and confinement in anisotropy-modulated multiferroic heterostructures. *Phys. Rev. Appl.* **8**, 014020 (2017).

37  Wong, K. L. *et al.* Unidirectional propagation of magnetostatic surface spin waves at a magnetic film surface. *Appl. Phys. Lett.* **105**, 232403 (2014).

38  Liu, T. *et al.* Ferromagnetic resonance of sputtered yttrium iron garnet nanometer films. *J. Appl. Phys.* **115**, 87-90 (2014).

39  Madami, M., Gubbiotti, G., Tacchi, S. & Carlotti, G. Application of microfocused Brillouin light scattering to the study of spin waves in low dimensional magnetic systems. *Solid State Phys.* **63**, 79-150 (2012).

40  Topp, J., Heitmann, D., Kostylev, M. P. & Grundler, D. Making a reconfigurable artificial crystal by ordering bistable magnetic nanowires. *Phys. Rev. Lett.* **104**, 207205 (2010).

41  Kalinikos, B. A. & Slavin, A. N. Theory of dipole-exchange spin wave spectrum for ferromagnetic films with mixed exchange boundary conditions. *J. Phys. C Solid State Phys.* **19**, 7013–7033 (1986).

42  Klingler, S. *et al.* Spin-torque excitation of perpendicular standing spin waves in coupled YIG/Co heterostructures. *Phys. Rev. Lett.* **120**, 127201 (2018).

43  Garcia-Sanchez, F., Borys, P., Vansteenkiste, A., Kim, J. V. & Stamps, R. L. Nonreciprocal spin-wave channeling along textures driven by the Dzyaloshinskii-Moriya interaction. *Phys. Rev. B* **89**, 224408 (2014).

44  Zhang, X., Zou, C. L., Jiang, L. & Tang, H. X. Strongly coupled magnons and cavity microwave photons. *Phys. Rev. Lett.* **113**, 156401 (2014).

45  M. Donahue, D. Porter, OOMMF User's Guide, Version 1.0, National Institute of Standards and Technology, Gaithersburg, MD, interagency report nistir 6376 Edition (Sept 1999). URL http://math.nist.gov/oommf.

46  Yu, H. *et al.* Omnidirectional spin-wave nanograting coupler. *Nat. Commun.* **4**, 2702 (2013).

47  Au, Y. *et al.* Resonant microwave-to-spin-wave transducer. *Appl. Phys. Lett.* **100**, 182404 (2012).

48  Wang, Q. *et al.* Reconfigurable nanoscale spin-wave directional coupler. *Sci. Adv.* **4**, 1701517 (2018).

49  Rodríguez-Fortuño, F. J. *et al.* Near-field interference for the unidirectional excitation of electromagnetic guided modes. *Science* **340**, 328-330 (2013).

50  Hämäläinen, S. J., Madami, M., Qin, H., Gubbiotti, G. & van Dijken, S. Control of spin-wave transmission by a programmable domain wall. *Nat. Commun.* **9**, 4853 (2018).

**Author contributions**